\documentclass{PoS}
\usepackage{colordvi}
\usepackage{multirow}
\usepackage{amsmath,amsfonts,epsfig,cite,graphics}
\usepackage{amssymb,bbm}
\usepackage{mathrsfs}
\usepackage{amsbsy}
\usepackage{latexsym}
\usepackage{graphicx}
\usepackage{psfrag}
\usepackage{macros_static}

\newcommand{\srdg}{Schr\"odinger }

\title{Cutoff effects of heavy quark vacuum polarization at one-loop order. \\ \hspace{\stretch{0}} { \hspace{\stretch{1}} {\large{\rm {DESY 11-181}}}} \\ \vspace{-0.25cm} \hspace{\stretch{1}} {\vspace{-0.75cm}\large{\rm {SFB/CPP-11-58}}}}
\ShortTitle{Cutoff effects of heavy quark vacuum polarization at one-loop order.}
\author{Andreas Athenodorou \\ NIC, DESY, Platanenallee 6\\
        15738 Zeuthen, Germany\\
        E-mail: \email{Andreas.Athenodorou@desy.de}}

\abstract{
The charm-quark mass is typically not so far from the cutoff $a^{-1}$ in lattice simulations. Its determinant may then potentially introduce large cutoff effects. We choose the O($a$)-improved Wilson formulation and compute the vacuum polarization effects in two rather different observables at one-loop order. One is the quark-antiquark static force and the other the \srdg functional coupling; in addition we investigate two more quantities resulting from the latter. In all the cases the lattice artifacts due to the charm-quark are small when compared to the gluonic effects. This indicates that the inclusion of charm-quarks in dynamical fermion simulations is typically not a problem. %However, non-perturbative investigations of lattice artifacts have to be carried out before a stronger statement can be made.
}

\PACS{12.38.Gc, 14.65.Dw, 12.38.Bx, 11.10.Gh \hspace{\stretch{1}} }

\FullConference{The XXIX International Symposium on Lattice Field Theory\\
         July 10-16 2011\\
         Squaw Valley, Lake Tahoe, California, USA}

\begin{document}
\vskip -0.25cm
\section{Introduction}
\label{Introduction}
\vspace{-0.35cm}  
Some lattice groups have already started investing effort in studies involving charm-quark vacuum polarization in Lattice QCD simulations~\cite{lat10:gregorio}. The main goal of these is to exclude noticeable corrections due to charm-quark loops. One, for instance, could think of the non-perturbative computation of the quark masses and the $\Lambda$-parameter in QCD%; this can be performed by the step-scaling method. 
. If in a such a computation we can only use the three-flavor theory then the connection to the four-flavor theory can be done just perturbatively. The ALPHA-collaboration has already started computations in $N_f=4$ using as a first observable the running-coupling in the massless \srdg functional scheme\cite{alpha:nf4}. However, in order to set the scale, observables in the massive theory have to be computed. A typical lattice spacing is $a \sim (2 {\rm GeV})^{-1}$. Hence, the question of whether such computations are feasible is arising. 

Our main worry occurs from the fact that in this case the charm-quark mass in lattice units is as large as $am_c = 1/2$. The main problem results from the findings of \cite{zastat:pap2,hqet:pap2} according to which in the valence-quark sector of the $\rmO(a)$-improved theory, although for masses up to $am_c = 1/2$ the cutoff effects are sizeable and obey approximately a quadratic behaviour in $a$, above this mass the Symanzik analysis and improvement of cutoff effects breaks down. %Hence, for masses larger than $am_c = 1/2$ Symanzik $\rmO(a)$-improvement ceases to be useful. 

The purpose of our work is to investigate how big the cutoff effects in charm-quark vacuum polarization effects are. We, therefore, expand perturbatively a few observables in the renormalized coupling and investigate the dependence of the first non-trivial expansion coefficient on the lattice spacing and the quark mass. We make use of our previous work with H. Panagopoulos~\cite{pertforce:andreasharis} and that by Rainer Sommer and S. Sint~\cite{pert:1loop} in order to extract the lattice cutoff effects due to fermions in the ${\rm q{\overline q}}$-force $F(r)$ and the \srdg functional coupling respectively. Using the latter we defined three different observables, namely the step-scaling function $\Sigma$ and the renormalized quantities $\vbar$ and $\rho$; for these quantities we extract the cutoff effects. This script is a compact version of Ref.~\cite{alpha:potsf}.
\vspace{-0.65cm}  
\section{Lattice Formulation}
\label{latticeformulation}
\vspace{-0.35cm}  
We begin by commenting briefly on the most important features of the O($a$)-improved theory. All the relevant information is described in detail in~\Ref{impr:pap1}. According to Wilson's lattice regularization the total action $S=S_g+S_f$ is given by:
\vspace{-0.25cm}  
\begin{eqnarray}
  S_g[U] = \frac{1}{g_0^2}\sum_{p}w(p)\,\tr\,\{1-U(p)\} \ \ \ \ {\rm and} \ \ \ \ S_f[U,\bar\psi,\psi] = a^4\sum_{x}\sum_{i=1}^{\nf}\bar\psi_i(D+\mibare)\psi_i , \label{e:SgSf}
\end{eqnarray} 
\vskip -0.25cm
\noindent
where the gluonic part $S_g$ is expressed as a sum over all oriented plaquettes $p$ with $U(p)$ the path-ordered product of the gauge fields around $p$. We consider an infinite lattice and the weight factors $w(p)$ are set to unity for now. The Dirac operator $D$ reads:
\vspace{-0.25cm}  
\begin{equation}
  D = \frac12 \sum_{\mu=0}^3
  \{\gamma_\mu(\nabla_\mu^\ast+\nabla_\mu^{})- a\nabla_\mu^\ast\nabla_\mu^{}\}
 + \csw\,\frac{ia}{4}\sum_{\mu,\nu=0}^3
 \,\sigma_{\mu\nu}\hat F_{\mu\nu}, \label{e:Dlat}
\end{equation}
\vskip -0.25cm
\noindent
with $\nabla_\mu$ and $\nabla_\mu^{\ast}$ the forward and backward covariant derivatives respectively and $\csw=1+\rmO\left(g_0^2\right)$. One can perform a systematic investigation of lattice cutoff effects after renormalizing the theory; hence, the cutoff effects depend upon the renormalization conditions. We, therefore, first use a massless renormalization scheme with scale $\mu$ and then a massive scheme at scale $\mu$ in order to investigate cutoff effects in lattice perturbation theory.

At one-loop order, the renormalized coupling and quark masses\footnote{At higher orders in perturbation theory, the sructure of the renormalization and the improvement is more complicated when all quark masses are considered. More information can be found in~\Ref{impr:nondeg}.} are given by:
\vspace{-0.25cm}
\bes
  \gbar^2(\mu)=\gtilde^2\zg \left(\gtilde^2,a\mu\right)
  \quad {\rm and} \quad
  \mir(\mu)=\mqitilde + \rmO\left(g_0^2\right),
\ees
\vskip -0.25cm
\noindent
respectively in terms of the improved bare coupling and improved bare mass \cite{pert:1loop,impr:pap1} :
\vspace{-0.25cm}
\bes
   \gtilde^2=g_0^2\left(1+{0.01200(2)}g_0^2 a\sum^{N_f}_{i=1} \mibare\right) + \rmO\left(g_0^6\right) 
   \quad {\rm and} \quad
  \mqitilde=\mibare\left(1-\frac12a\mibare\right)\,.
\ees
\vskip -0.15cm
\noindent
For a complete ${\rm O}(a)$-improvement \`a la Symanzik one has to use the modified-bare coupling and modified-quark-mass in a massless renormalization scheme.
\vspace{-0.35cm}
\section{A systematic way of investigating the cutoff effects}
\vspace{-0.25cm}
\label{cutoffeffects}
It is usually normal to consider observables the perturbative expansion of which in the minimal-subtraction scheme $\gbarMSbar$ is known in the continuum. We, therefore, renormalize in the $\msbar$ scheme. We do so by first moving from the modified-bare coupling to the lattice-minimal-subtraction~(lat) scheme and then to the $\msbar$ scheme. The lattice-minimal-subtraction scheme is defined in such a way so that we subtract the logarithmic divergences in $a$ order by order in perturbation theory. For the perturbative order we are interested in:
\vspace{-0.25cm}
\begin{eqnarray}
  \glat^2(\mu)=\gtilde^2\zlat\left(\gtilde^2,a\mu\right) \quad {\rm with} \quad \zlat\left(\gtilde^2,a\mu\right) = 1 -2b_0 \gtilde^2 \log(a\mu) + \rmO\left(\gtilde^4\right)\,,
\end{eqnarray}
\vskip -0.25cm
\noindent
where {\small ${b_0=\left( {11 N_c}/{3}   - {2N_f}/{3} \right) / (4 \pi)^2}$}. We use observables $O$ that depend on a single length scale, let us say, $r$ and on the masses $\mir$. These have the following perturbative expansion:
\vspace{-0.25cm}
\bes
  O = g_0^2 + O^{(1)}(\vecz,a/r)\,g_0^4 + \ldots\,, 
  \label{e:Oexpansion1}
\ees
\vskip -0.25cm
\noindent
where $\vecz = (z_1,\ldots,z_{\nf})$ and $z_i=\mir\cdot r$. After moving from the bare to the modified-bare-coupling and then to the lattice-minimal-subtraction scheme, the expansion of the observable takes the form:
\vspace{-0.75cm}
\bes
   O = \tilde O_\mrm{cont}\left(r\mu,\vecz,\glat^2(\mu)\right)
       \left(1 + \tilde\delta_O(r\mu,\vecz,\glat^2(\mu),a/r)\right),
       \label{e:Oexpression1}
\ees 
\vskip -0.25cm
\noindent
written as a continuum part and a lattice part $\tilde\delta_O(r\mu,\vecz,\glat^2(\mu),a/r)$. The latter is expressed as:
\vspace{-0.25cm}
\bes
   \tilde\delta_O\left(r\mu,\vecz,\glat^2(\mu),a/r\right) =  
   \tilde\delta_O^{(0)}(r\mu,\vecz,a/r) +
   \tilde\delta_O^{(1)}(r\mu,\vecz,a/r) \glat^2(\mu) + \ldots \,.
\ees
\vskip -0.25cm
\noindent
From the expression (\ref{e:Oexpression1}) $\tilde\delta_O(r\mu,\vecz,\glat^2(\mu),a/r = 0) =0$. We can now move to the $\msbar$ by applying the finite scheme transformation  \cite{pert:2loopLW,pert:1loop,Bode:2001uz}:
\vspace{-0.25cm}
\bes
   \glat^2(\mu) &=& \gbarMSbar^2 (\mu)  
   - \frac{c_{1}^{\rm lat,\MSbar}}{4 \pi} \gbarMSbar^4 (\mu) 
   + \rmO \left( \gbarMSbar^6  (\mu)    \right),
\label{e:defc1}
 \\
 c_1^{\rm lat,\MSbar}&=&  -\frac{ \pi }{2 \nc} + 2.135730074078457(2) \nc -0.39574962(2) \nf \,.\quad 
\ees
\vskip -0.25cm
\noindent
The expansion of the relative cutoff effects is written as:
\vspace{-0.25cm}
\bes
 \delta_O\left(r\mu,\vecz,\gbarMSbar^2(\mu),a/r\right) &\equiv&
 \left.{O - O_\mrm{cont} \over O_\mrm{cont}}\right|_{\gbarMSbar,\mir} = \delta_O^{(1)}(r\mu,\vecz,a/r) \gbarMSbar^2(\mu) + \ldots \,.
 \label{e:arteexp}
\ees
\vskip -0.25cm
\noindent
Since the theory is renormalized by minimal-subtraction, these lattice artifacts are intrinsically perturbative and have no non-perturbative extension. Nevertheless, one can use combinations of $\delta_O^{(i)}$ from different observables in order to obtain the expansion coefficients of the non-perturbative cutoff effects. In the longer write-up~\cite{alpha:potsf} we demonstrate how by renormalizing in%, and using as an observable, 
  \ the \srdg functional coupling $\gbarSF^2(L)$ for massless quarks we obtain the non-perturbative lattice artifacts at small couplings. Their fermionic part is approximately the same as that of Eq.~(\ref{e:arteexp}) i.e. $\sim \delta_O^{(1,f)}(r\mu,\vecz,a/r)$. 

%\newpage
However, it is more convenient to introduce a massive renormalization scheme for non-perturbative computations in QCD with a charm-quark. This is due to the fact that the mass of the charm-quark is larger than the typical QCD scale, thus, it has reduced vacuum polarization effects which is most efficiently implemented by using a finite mass renormalisation scheme.

We can define a massive scheme with scale $\mu=1/r_0$ through a particular observable $O_0$:
\vspace{-0.25cm}
\bes
  \gbar^2_{m}(\mu,\vecmr) \equiv O_0 = g_0^2 + O_0^{(1)}(\vecz,a/r)\,g_0^4 + \ldots\,.
  \label{e:Oexpansion} 
\ees
\vskip -0.25cm
\noindent
It is straightforward to show~\cite{alpha:potsf} that for an observable $O$ at a different length scale $r$ the relative artifacts in the massive scheme are given as a combination of those in the massless:
\vspace{-0.25cm}
\bes
\label{e:deltamassive}
\delta_{O_{\rm m}} \hspace{-1.0mm} \left( \hspace{-0.5mm} r/r_0,\vecz,\gbarm^2(\mu,\vecmr\hspace{-0.5mm}),a/r \hspace{-0.5mm} \right)
  \equiv
   \left[\hspace{-0.5mm} \delta_O^{(1)} \hspace{-0.5mm} ( \hspace{-0.5mm} r\mu,r\vecmr,a/r \hspace{-0.5mm} ) \hspace{-0.5mm} - \hspace{-0.5mm}
         \delta_{O_0}^{(1)} \hspace{-0.5mm} ( \hspace{-0.5mm} r\mu,r_0\vecmr,a/r_0 \hspace{-0.5mm})\hspace{-0.5mm} \right]\, \hspace{-1.0mm}
   \gbarm^2(\mu,\vecmr \hspace{-0.5mm}) \hspace{-0.5mm} + \hspace{-0.5mm} \ldots \,.
\ees
\vskip -0.25cm
\vspace{-0.45cm}
\section{The ${\rm q {\bar q}}$-static-Force.}
\label{Force}
\vspace{-0.25cm}
We start our investigation by considering the force $F(r)= \frac{d}{dr} V(r)$ between static quarks. If one uses on-axis potentials as we do here, the most natural choice to define the derivative would be: $F_\mrm{naive}(r_\mrm{naive}) = \frac1a \left[ V(r,0,0)-V(r-a,0,0) \right]$ with $r_\mrm{naive}=r-\frac{a}{2}$. However, the force can also be defined in terms of an improved ${\rm q {\bar q}}$-separation $r_I$ according to which the force has no effects at tree-level order in perturbation theory~\cite{pot:intermed}, thus, {\small $F_{\rm tree}(r_I)={\cf}/{4\pi \rI^2}$} with  {\small $\cf = (N_c^2-1)/2N_c$}. The improved separations $r_I$ for on-axis potentials were calculated in~\cite{pot:intermed}. It is worth mentioning that the non-perturbative force in the pure gauge theory when defined in terms of $r_I$ has much smaller lattice artifacts compared to $r_{\rm naive}$. The results presented below are expressed in terms of $r_I$ while we have in addition looked at all quantities for $r_{\rm naive}$ with no any worth reporting changes; hence, from here on we have $r=r_I$. 

The one-loop corrections to the force are calculated as described in~\cite{pertforce:andreasharis}. The expression of the force in the  $\msbar$ scheme after setting the renormalization scale to the natural choice $\mu=1/r$ is given by:
\vspace{-0.25cm}
\bes
  F = {\cf \alphaMSbar(1/r)  \over r^2} 
      \left\{1 + f_1(\vecz,a/r) \alphaMSbar(1/r) 
               + \rmO(\alphaMSbar^2) \right\}\,. 
\ees
\vskip -0.25cm
\noindent
The term $f_1$ is split into a gluonic~($g$) and fermionic~($f$) contributions such as:
\vspace{-0.25cm}
\bes
   f_1&=& f_{1,g}(a/r) + \sum^{N_f}_{i=1} f_{1,f} (z_i,a/r)\,,\quad 
\ees
\vskip -0.25cm
\noindent
with continuum expressions~\cite{pot:Hoang} ($\gamma_E=0.57721566\dots$):
\vspace{-0.25cm}
\bes
   f_{1,g}(0) &=& {\nc \over \pi} \left[  - \frac{35}{36} +\frac{11}{6} \gamma_{E} \right], 
   \label{e:cng} \\
   f_{1,f} (z,0) & = & \frac{1}{2\pi} \left[ \frac{1}{3} {\rm log}(z^2)
                    + \frac{2}{3} \int_{1}^{\infty} \rmd x 
    \frac{1}{x^2} \sqrt{x^2-1}\left(1+\frac{1}{2x^2}  \right) 
                  \left( 1+2 z x  \right)e^{-2zx}   \right] \,.
   \label{e:cnf}
\ees 
\vskip -0.25cm
\noindent
The relative lattice artifacts of the force are defined as:
\vspace{-0.25cm}
\bes
  \left.{F - F_\mrm{cont} \over F_\mrm{cont}}\right|_{\gbarMSbar, \mir} = 
  \delta_F^{(1)}(\vecz,a/r) \gbarMSbar^2(1/r)+\dots, 
\ees
\vskip -0.25cm
\noindent
with\footnote{The pre-factor $4 \pi$ on $\delta_F^{(1)}$ appears because we multiply with $\alpha=\gbar^2/(4\pi)$.}:
\vspace{-0.25cm}
\bes
  4 \pi \delta_F^{(1)}(\vecz,a/r) = 4 \pi \delta_F^{(1,g)}(a/r) +
                            4 \pi \sum_{i=1}^{\nf} \delta_F^{(1,f)}(z_i,a/r), %\ \
   %4 \pi \delta_F^{(1,f)}(z,a/r) = f_{1,f} (z,a/r) - f_{1,f} (z,0)\,.
\ees
\vskip -0.25cm
\noindent
The gluonic contribution to the force at finite $a/r$ is not our main concern. However, we need it for a complete picture on the size of the lattice cutoff effects. We, therefore, provide numbers for $\delta_F^{(1,g)}(a/r)$ extracted from~\Ref{pot:pertBB}. Concerning the fermionic piece, we perform a re-evaluation of $\delta_F^{(1,f)}(z,a/r)$ for the relevant bare-masses according to~\Ref{pertforce:andreasharis}. For more details on the procedure adopted to extract $\delta_F^{(1,f)}(z,a/r)$ see~\Ref{alpha:potsf}.

\begin{figure}[tb!]
\vspace{-0.65cm}
\centerline{\scalebox{0.7}{\input{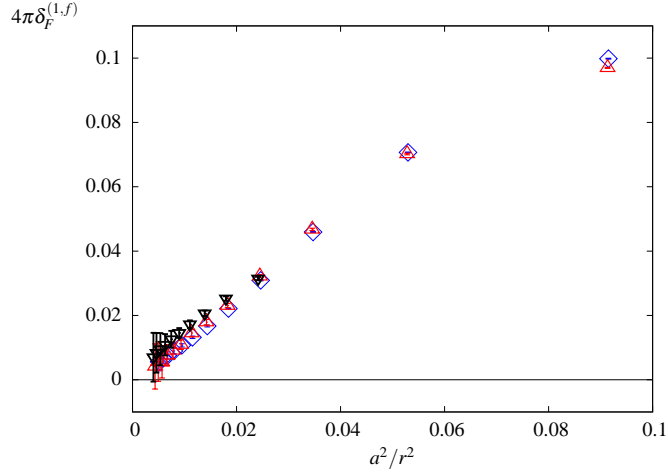}}}
\vspace*{-3mm}
\caption[]{\label{f:df1f}
The fermionic cutoff effects $4\pi \delta_F^{(1,f)}(z,a/r)$ for $z=0$~({\LARGE\Blue{$\diamond$}}), 
$z=1$~({\small \Red{$\triangle$}}) and 
$z=3$~($\triangledown$).}
\vspace{-0.45cm}
\end{figure}

In order to obtain a first glimpse on the relevant size of the lattice artifacts let us view the gluonic part $4 \pi \delta_F^{(1,g)}(a/r)$. For $r/a=2.277$, $3.312$ and $4.319$, $4\pi \delta_F^{(1,g)}(a/r)  = -0.232(6)$, $-0.190(19)$ and $-0.151(42)$ respectively. In Fig.~\ref{f:df1f} we display results for the fermionic contribution to the lattice artifacts $4 \pi \delta_F^{(1,f)}(z,a/r)$. The scale of the y-axis is about a factor two smaller than the pure gauge artifacts. Hence, the absolute value of the fermionic piece of the cutoff effects for a single flavor is much less than the gluonic. Concerning the mass dependence, we see that the fermionic cutoff effects depend very little on the mass.
%%%%%%%%%%%%%%%%%%%%%%%%%%%%%%%%%%%%%%%%%%%%%%%%%%%%%%%%%%%%
%%%%%%%%%%%%%%%%%%%%%%%%%%%%%%%%%%%%%%%%%%%%%%%%%%%%%%%%%%%%
\vspace{-0.35cm}
\section{\srdg Functional}
\label{schrodingerfunctional}
\vspace{-0.35cm}
We call \srdg functional the field theory in a finite space-time with volume $L^4$, Dirichlet boundary conditions along the time direction and periodic in space up to a phase $\theta$ for the quark fields. According to Monte-Carlo simulations~\cite{pert:1loop} the advantageous value of $\theta$ is $\theta=\pi/5$ while $\theta=0$ is the natural, and more aesthetically appealing, alternative. The action has the form~(\ref{e:SgSf}) with all fields being zero outside $0 \le x_0 \le L$, the gauge fields having fixed values at the boundaries $x_0=0$ and $L_0=0$ and the fermionic fields being zero at these boundaries. For a complete ${\rm O}(a)$-improvement one has to modify the weight factor of the timelike plaquettes attached to the boundaries $w(p)= \ct=1-[0.08900 + 0.019141\nf \pm 0.00005]g_0^2 +\ldots$; for more details we refer to \Refs{impr:pap1,SF:LNWW,pert:1loop}.

%The importance of the \srdg functional functional is mostly due to the fact that we can simulate it for massless quarks and define and calculate precisely a running coupling in a Monte-Carlo simulation. 
For our perturbative calculation we use the definition of the coupling appearing in \Refs{SF:LNWW,pert:1loop}. The \srdg functional coupling depends upon $L$ playing the r\^ole of the inverse renormalization scale, the phase $\theta$ and the dimensionless parameter $\nu$ which appears on the fixed boundary gauge fields and is usually set to zero. In a non-perturbative computation of the running of the coupling the main object needed is the step-scaling-function defined as:
\vspace{-0.25cm}
\bes
   \Sigma(u,\vecz,a/L) \equiv \gbarSF^2(2L,2z)|_{\gbarSF^2(L,\vecz)=u, \mir L=z_i}
    = u + \Sigma_1(\vecz,a/L)u^2 + \ldots \,.
\ees
\vskip -0.25cm
\noindent
However, in order to have a more general picture we can also look at other quantities such as:
\vspace{-0.25cm}
\bes
   \Omega(u,\vecz,a/L) \equiv \vbar(L,\vecz)|_{\gbarSF^2(L,\vecz)=u, \mir L=z_i} 
   = \vbar_1(\vecz,a/L)+\rmO(u)\,,
\ees
\vskip -0.25cm
\noindent
with the quantity $\vbar$ defined explicitly by:
\vspace{-0.25cm}
\bes
  {1\over \gbar^2_\nu(L,\nu,\vecz)} = {1\over\gbarSF^2(L,\vecz)}-\nu\, \vbar(L,\vecz)\,,
\ees
\vskip -0.25cm
\noindent
and
\vspace{-0.25cm}
\bes
   \rho(u,\vecz,a/L) \equiv 
   \frac{\gbarSF^2(L,\vecz) - \gbarSF^2(L,\veczero)}{\gbarSF^2(L,\veczero)} \biggr\vert_{\gbarSF^2(L,0)=u, \mir L=z_i} = 
   \rho_1(\vecz,a/L)u + \rmO(u^2)\,.
\ees
\vskip -0.25cm
\noindent
For the step-scaling function ${\Sigma}$, we consider the relative lattice artifacts:
\vspace{-0.25cm}
\bes
   \delta_\Sigma(u,\vecz,a/L) = 
   {\Sigma(u,\vecz,a/L) - \Sigma(u,\vecz,0) \over \Sigma(u,\vecz,0)} 
   = \delta_\Sigma^{(1)}(\vecz,a/L) u + \ldots \,,
\ees
\vskip -0.25cm
\noindent
while for both $\rho$ and ${\vbar}$ the absolute artifacts:
\vspace{-0.25cm}
\bes
   \Delta_{\vbar}(u,\vecz,a/L) &\equiv& \Omega(u,\vecz,a/L) - \Omega(u,\vecz,0)
   =
   \Delta_{\vbar}^{(1)}(\vecz,a/L)  + \rmO(u) \,,  
\\
   \Delta_\rho(u,\vecz,a/L) &\equiv& \rho(u,\vecz,a/L) - \rho(u,\vecz,0)
   =
   \Delta_\rho^{(1)}(\vecz,a/L)u  + \rmO(u^2) \,.
\ees
\vskip -0.25cm
\noindent
The leading perturbative terms $\delta_\Sigma^{(1)}(\vecz,a/L)$, $\Delta_{\vbar}^{(1)}(\vecz,a/L)$ and $\Delta_\rho^{(1)}(\vecz,a/L)$ can be decomposed into gluonic and fermionic pieces:
\vspace{-0.25cm}
\bes
 \delta_\Sigma^{(1)}(\vecz,a/L)  
   &=& \delta_\Sigma^{(1,g)}(\vecz,a/L) + {\rm \sum_{i=1}^{\nf}} \delta_\Sigma^{(1,f)}(z_i,a/L),
   \\
 \Delta_{\vbar}^{(1)}(\vecz,a/L)  
   &=& \Delta_{\vbar}^{(1,g)}(a/L) + \sum_{i=1}^{\nf}\Delta_{\vbar}^{(1,f)}(z_i,a/L),
   \\
 \Delta_{\rho}^{(1)}(\vecz,a/L)  
   &=& \sum_{i=1}^{\nf}\Delta_\rho^{(1,f)}(z_i,a/L)\,.
\ees
\vskip -0.25cm
\noindent
\begin{figure}[tb!]
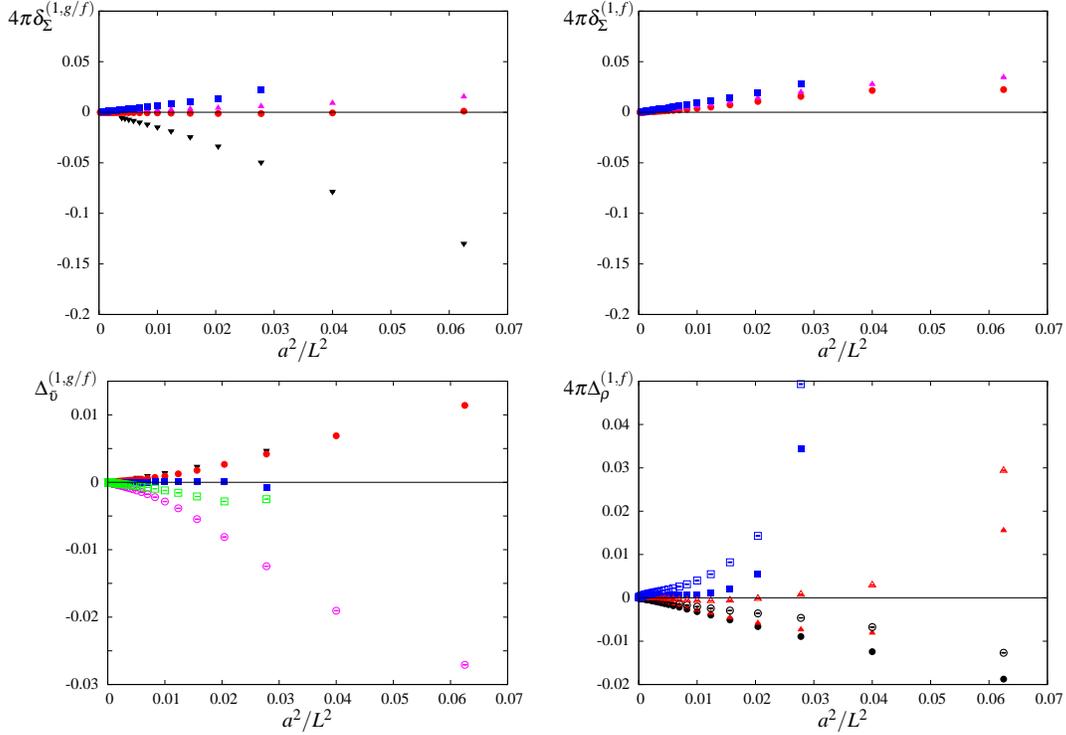

\vskip -0.65cm
\centerline{\scalebox{1}{\scalebox{0.55}{\input{Stepscaling1.tex}}  \hspace{0.0cm}  \scalebox{0.55}{\input{Stepscaling2.tex}}}}
\centerline{\scalebox{1}{\scalebox{0.55}{\input{upsilon.tex}}  \hspace{0.0cm}  \scalebox{0.55}{\input{rho.tex}}}}
\vspace{-0.25cm}
\caption{{\underline{Upper-Left:} The cutoff effects $4\pi \delta_\Sigma^{(1,g)}(a/L)$ ({\large $\blacktriangledown$}) and  $4\pi  \delta_\Sigma^{(1,f)} (z,a/L)$ for $z=0$~({\LARGE \Red{$\bullet$}}), $z~=~1$~({\large\Magenta{$\blacktriangle$}}) and $z=3$~({\Blue{$\blacksquare$}}) extracted for $\theta=\pi/5$.}  {\underline{Upper-Right:} $4\pi  \delta_\Sigma^{(1,f)} (z,a/L)$ for $z=0$~({\LARGE\Red{$\bullet$}}), $z=1$~({\large\Magenta{$\blacktriangle$}}) and $z=3$~({\Blue{$\blacksquare$}}) extracted for $\theta=0$.} {\underline{Lower-Left:} The cutoff effects 
$\Delta_{\vbar}^{(1,g)}(a/L)$ ({\small $\blacktriangledown$}) and $\Delta_{\vbar}^{(1,f)}(z,a/L)$  for $z=0$ ({\LARGE \Red{$\bullet$}}~({\scriptsize\Magenta{$\bigcirc$}})) and
$z=3$ ({\Blue{$\blacksquare$}} ({\Green{$\square$}})) extracted for $\theta=\pi/5 \ (\theta=0)$.} {\underline{Lower-Right:} The cutoff effects $4\pi  \Delta_{\rho}^{(1,f)}(z,a/L)$ for $z=1$~({\LARGE $\bullet$} ({\scriptsize $\bigcirc$})), $z=2$~({\large\Red{$\blacktriangle$}} ({\small\Red{$\triangle$}})) 
and $z=3$~({\Blue{$\blacksquare$}} ({\Blue{$\square$}})) extracted for $\theta=\pi/5$ ($\theta=0$).}} 
\label{sfunctional}
\vspace{-0.35cm}
\end{figure}

In the upper-left and upper-right plots of Fig.~\ref{sfunctional} we show the artifact $4\pi  \delta_\Sigma^{(1,f)} (z,a/L)$ for $\theta=\pi/5$ and $\theta=0$ respectively. The cutoff effects for an individual ${\rm O}(a)$-improved fermion are much smaller than those for the gluonic part. They do appear to grow with $z$, however, not very much. Of course one should always bear in mind that for more than one fermion flavor the cutoff effects add up accordingly. 

If we now move to the cutoff effects of the ${\vbar}$ which appears in the lower-left plot of Fig.~\ref{sfunctional}, we observe that are bigger compared to the step-scaling function (given that the overall magnitude is $\vbar_1 \approx 0.1$). In contrast to what we observed before, here, the cutoff effects decrease with the mass.

Finally, in the lower-right plot of Fig.~\ref{sfunctional} we present the absolute lattice artifacts in $\rho$. These have to be compared to the continnum values of $4\pi {\rho}(z,0) $ which range from 0.095 (0.086) for $z=1$ up to 0.188 (0.170) for $z=3$ at $\theta=\pi/5$ ($\theta=0$).

\vspace{-0.35cm}
\section{Conclusion}
\label{conclusions}
\vspace{-0.35cm}
We have investigated the cutoff effects of several observables in two rather different setups from including an ${\rm O}(a)$-improved Wilson charm-quark at one-loop order in lattice perturbation theory. We always restrict ourselves to $am_c<1/2$. We observed that these are comparable to the cutoff effects of the pure gluonic part; for instance see the lower-left plot of Fig.~\ref{sfunctional}. As a matter of fact, in some cases the fermionic pieces of the cutoff effects are even smaller than their pure gluonic ones. This is demonstrated clearly in the upper-left plot of Fig.~\ref{sfunctional}. The reason of studying these was to check whether they become very large for very-massive quarks and more specifically for masses associated with the charm-quark. In contrast to our expectations of seeing large cutoff effects, we observe that they do not grow much with the mass and sometimes even decrease for larger masses. The overall conclusion is that the lattice artifacts remain small as we increase the mass and, thus, the inclusion of charm-quarks in dynamical fermion simulations does not seem to be a problem with the current available lattice spacings. However, non-perturbative investigations of lattice artifacts have to be carried out.
\vspace{-0.35cm}
\section*{Acknowledgements}
\vspace{-0.35cm}
We \hspace{\stretch{1}} would \hspace{\stretch{1}} like \hspace{\stretch{1}} to \hspace{\stretch{1}} thank \hspace{\stretch{1}} Rainer Sommer, \hspace{\stretch{1}} Haris Panagopoulos \hspace{\stretch{1}} and \hspace{\stretch{1}} Ulli \hspace{\stretch{1}} Wolff \hspace{\stretch{1}} for \hspace{\stretch{1}} usefull \hspace{\stretch{1}} discussions \hspace{\stretch{1}} and \hspace{\stretch{1}} Matthias \hspace{\stretch{1}} Steinhauser \hspace{\stretch{1}} for \hspace{\stretch{1}} pointing \hspace{\stretch{1}} out \hspace{\stretch{1}} Eq.~(\ref{e:cnf}). \hspace{\stretch{1}} The numerical computations were carried out on the compute farm at DESY, Zeuthen. This work is supported by the SFB/TR 09  of the Deutsche Forschungsgemeinschaft.  \hspace{\stretch{1}}

\end{document}